\documentclass[showpacs,preprintnumbers,amsmath,amssymb,superscriptaddress]{revtex4}

\usepackage{bm}
\usepackage{graphicx}
\begin{document}

%-----------------------------------------------------------------
\title{Temporary mirror symmetry breaking and chiral excursions in
  open and closed systems}
%-----------------------------------------------------------------
\author{Celia Blanco}
\email{blancodtc@inta.es} \affiliation{Centro de Astrobiolog\'{\i}a
(CSIC-INTA), Carretera Ajalvir Kil\'{o}metro 4, 28850 Torrej\'{o}n
de Ardoz, Madrid, Spain}
\author{Michael Stich}
\email{stich@inta.es} \affiliation{Centro de Astrobiolog\'{\i}a
(CSIC-INTA), Carretera Ajalvir Kil\'{o}metro 4, 28850 Torrej\'{o}n
de Ardoz, Madrid, Spain}
\author{David Hochberg}
\email{hochbergd@inta.es} \affiliation{Centro de Astrobiolog\'{\i}a
(CSIC-INTA), Carretera Ajalvir Kil\'{o}metro 4, 28850 Torrej\'{o}n
de Ardoz, Madrid, Spain}

\begin{abstract}
  The reversible Frank model is capable of amplifying the initial
  small statistical deviations from the idealized racemic composition.
  This temporary amplification can be interpreted as a chiral
  excursion in a dynamic phase space. It is well known that if the
  system is open to matter and energy exchange, a permanently chiral
  state can be reached asymptotically, while the final state is
  necessarily racemic if the system is closed. In this work, we
  combine phase space analysis, stability analysis and numerical
  simulations to study the initial chiral excursions and determine how
  they depend on whether the system is open, semi-open or closed.
\end{abstract}

\pacs{05.40.Ca, 11.30.Qc, 87.15.B-}
\date{\today}

\maketitle

%----------------------------------------------------------------
\section{\label{sec:intro} Introduction}
%---------------------------------------------------------------

The Frank model~\cite{Frank} has been extensively invoked to justify
theoretically the emergence of biological
homochirality~\cite{Mason,PKBCA}, and is usually analyzed as a
reaction network in open systems (matter and energy are exchanged
with the surroundings) composed of an irreversible enantioselective
autocatalysis coupled to an irreversible mutual inhibition reaction
between the product enantiomers. The model shows how homochirality
is achieved as a stationary state when the mutual inhibition product
(the heterodimer) is removed from the system and when the
concentration of the achiral substrate is held constant. By
contrast, for reversible transformations and when the mutual
inhibition product remains in the system, the \textit{final} stable
state can only be the racemic one. As a consequence, a
thermodynamically controlled spontaneous mirror symmetry breaking
(SMSB) cannot be expected to take place. In particular, SMSB is not
expected for reversible reactions taking place in systems closed to
matter and energy flow.

Nevertheless, as was recently demonstrated~\cite{CHMR} for systems
closed to matter flow, the Frank model is a prime candidate for the
fundamental reaction network necessary for reproducing the key
experimental features reported on absolute asymmetric synthesis in the
absence of any chiral polarization~\cite{Soai}.  Most importantly,
when reversible steps in all the reactions are allowed it is capable
of~\cite{CHMR} (i) amplification of the initially tiny statistical
enantiomeric excesses from $ee \sim 10^{-8}\%$ to practically 100\%,
leading to (ii) long duration chiral \textit{excursions} or chiral
pulses away from the racemic state at nearly 100\% $ee$, followed by,
(iii) the final approach to the stable racemic state for which $ee=0$,
i.e., mirror symmetry is recovered permanently. To understand this
temporary asymmetric amplification is important because the
racemization time scale can be much longer than that for the complete
conversion of the achiral substrate into enantiomers.

Long duration chiral excursions have also been reported recently in
closed chiral polymerization models with reversible
reactions~\cite{BH} where constraints implied by microreversibility
have been taken into account. These results are important because they
suggest that temporary spontaneous mirror symmetry breaking in
experimental chiral polymerization can take place, and with observable
and large chiral excesses without the need to introduce chiral
initiators~\cite{Lahav} or large initial chiral excesses~\cite{Hitz}.

The purpose of this Letter is to elucidate the nature of these chiral
excursions by combining the information provided by phase plane
portraits, numerical simulation and linear stability analysis. We
consider the Frank model, this being the most amenable to such types
of analysis and because it is the ``common denominator" of numerous
more elaborate theoretical models of SMSB~\cite{PKBCA}.

The reaction scheme consists of a straight non-catalyzed reaction
Eq.~(\ref{decay}), an enantioselective autocatalysis
Eq.~(\ref{autoLD2}), where A is a prechiral starting product, and L
and D are the two enantiomers of the chiral product. We also assume
reversible heterodimerization step in Eq.~(\ref{heterodimer}), where
LD is the achiral heterodimer. The $k_i$ denote the reaction rate
constants. In the following, we give the reaction steps in detail.

\noindent Production of chiral compound:
\begin{equation}\label{decay}
\textrm{A} \stackrel{k_1}{\rightleftharpoons \atop{k_{-1}}}
\textrm{L}, \qquad \textrm{A} \stackrel{k_1}{\rightleftharpoons
\atop{k_{-1}}} \textrm{D}.
\end{equation}
\noindent Autocatalytic amplification:
\begin{equation}\label{autoLD2}
\textrm{L} + \textrm{A} \stackrel{k_2}{\rightleftharpoons
\atop{k_{-2}}}\textrm{L} + \textrm{L}, \qquad \textrm{D} +
\textrm{A} \stackrel{k_2} {\rightleftharpoons \atop{k_{-2}}}
\textrm{D} + \textrm{D}.
\end{equation}
\noindent  Hetero-dimerization:
\begin{equation}\label{heterodimer}
\textrm{L} + \textrm{D} \stackrel{k_5}{\rightleftharpoons
\atop{k_{-5}}} \textrm{LD}.
\end{equation}
We assume the feasibility of the reverse reaction for all the steps.
Focusing our attention on chiral excursions, we make a careful
distinction between open, semi-open or fully closed systems. These
system constraints are crucial for determining both the intermediate
and the asymptotic final states of the chemical system.

%------------------------------------------------------------
\section{\label{sec:open} Open system}
%------------------------------------------------------------

%------------------------------------------------------------
\subsection{\label{sec:open1} Rate equations}
%------------------------------------------------------------

We first consider the original Frank scenario~\cite{Frank}. There,
steady and stable chiral states can be achieved, since the system is
permanently held out of equilibrium. See~\cite{RH} for more details.
An important question is, can the system support chiral excursions?
That is, pass through temporary chiral states before ending up in
the final racemic state? In the original Frank model there is an
incoming flow of achiral compound A and elimination of the
heterodimer LD from the system. A convenient way to account for the
inflow of achiral matter is to assume that the concentration of the
prechiral component $[A]$ is constant, and then we need not write
the corresponding kinetic equation for it. For the outflow the
heterodimer leaves the system at a rate $\bar \gamma$. We assume
that the heterodimer formation step is irreversible, and set
$k_{-5}=0$. Note, the elimination of LD from the system can actually
be neglected as long as the hetero-dimerization step is irreversible
\cite{PKBCA}. We retain this outflow however since it is needed to
obtain \textit{stationary} asymptotic values of all three
concentrations $[L],[D]$ and $[LD]$; see the fixed points below. So
with $[A]=const$ and replacing Eq. (\ref{heterodimer} ) by
\begin{eqnarray}\label{heterodim}
\textrm{L} + \textrm{D} &\stackrel{k_5}{\rightarrow}& \textrm{LD},\\
\textrm{LD} &\stackrel{\bar \gamma}{\rightarrow}& \emptyset ,
\end{eqnarray}
we obtain the rate equations
\begin{eqnarray}\label{Lopen}
\frac{d}{d t}[L] &=& k_1[A] + (k_2[A]-k_{-1})[L]-k_{-2}[L]^2 -
k_5[L][D], \\
\label{Dopen} \frac{d}{d t}[D] &=& k_1[A] + (k_2[A]-k_{-1})[D]
-k_{-2}[D]^2 - k_5[D][L], \\
%\label{A} \frac{d}{dt}[A] &=&
%-2k_1[A]-(k_2[A] - k_{-1})([L]+[D])+ k_{-2}([L]^2 + [D]^2), \\
\label{LDopen} \frac{d}{d t}[LD] &=& k_5[L][D] - \bar{\gamma} [LD].
\end{eqnarray}
The key variable throughout is the chiral polarization
\begin{equation}
\eta = \frac{[L]-[D]}{[L]+[D]},
\end{equation}
also called enantiomeric excess $ee$, which obeys $-1 \leq \eta \leq
1$ and which represents the order parameter for mirror symmetry
breaking.

In order to simplify the analysis, we define a dimensionless time
parameter $\tau = (k_2[A] - k_{-1})t$ and dimensionless concentrations
that scale as $[\tilde L] = k_5(k_2 [A] - k_{-1})^{-1} [L]$, $[\tilde
D] = k_5(k_2 [A] - k_{-1})^{-1} [D]$, $[\widetilde{LD}] = k_5(k_2 [A]
- k_{-1})^{-1} [LD]$.  It is convenient to define the sums and
differences of concentrations: $\chi = [\tilde L] + [\tilde D]$, $y =
[\tilde L] - [\tilde D]$, and for the heterodimer put $P =
[\widetilde{LD}]$. The chiral polarization $\eta= y/\chi$ remains
unchanged.

In terms of the new variables, Eqs.~(\ref{Lopen}-\ref{LDopen}) read
\begin{eqnarray}\label{chidotopen}
\frac{d \chi}{d \tau} &=& 2u + \chi
-\frac{1}{2}(g+1)\chi^2 -\frac{1}{2}(g-1)\chi^2 \eta^2,\\
\label{eeopen} \frac{d \eta}{d \tau} &=& \eta (1 - g \chi)
-\frac{\eta}{\chi}\left(\frac{d \chi}{d \tau}\right),\\ \label{Pdotopen}
\frac{d P}{d \tau} &=&\frac{1}{4}\chi^2 (1 - \eta^2) - \gamma P.
\end{eqnarray}
The dimensionless parameters appearing here are:
\begin{equation}\label{paramopen}
u = \frac{k_1 k_{5}[A]}{(k_2[A] - k_{-1})^2},\,\, g =\frac{k_{-2}}{k_5},
\,\, \gamma = \frac{\bar{\gamma}}{(k_2 [A] - k_{-1})}.
\end{equation}
The system is described by three equations
Eqs.~(\ref{chidotopen}-\ref{Pdotopen}). Since $P$ does not enter
into the equations for $\chi$ and $\eta$, the equations decouple and
the dynamical system to study is effectively two-dimensional and so
the appearance of SMSB cannot depend on whether the heterodimer is
removed from the system when $k_{-5}=0$, although the \textit{fixed
points} will (see Sec \ref{sec:openstab}).

%------------------------------------------------------------
\subsection{\label{sec:fullopennull}Phase plane and linear stability analysis}
%------------------------------------------------------------

\begin{figure}[h]
\begin{tabular}{cc}
\includegraphics[width=0.45\textwidth]{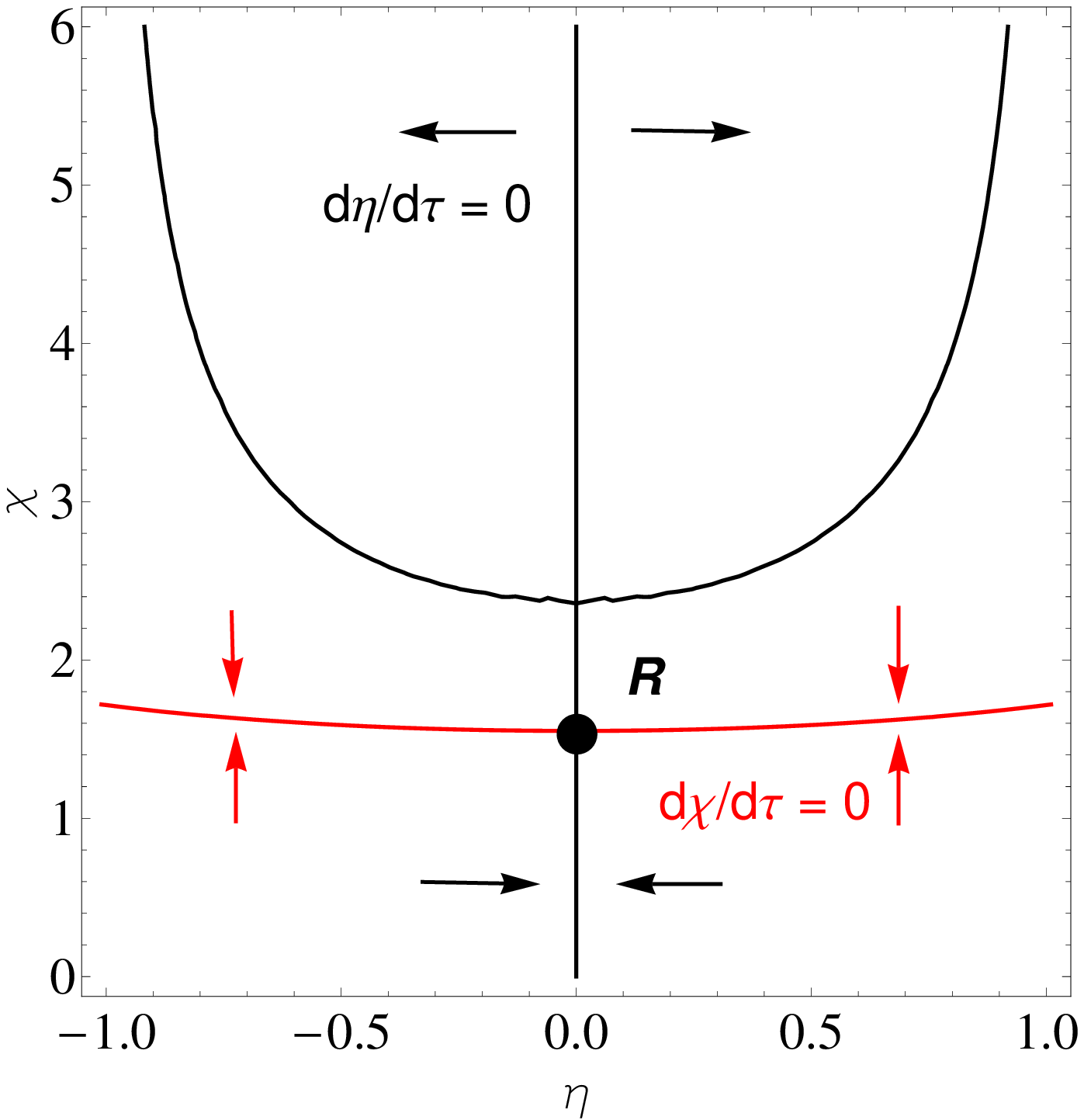}&
\includegraphics[width=0.45\textwidth]{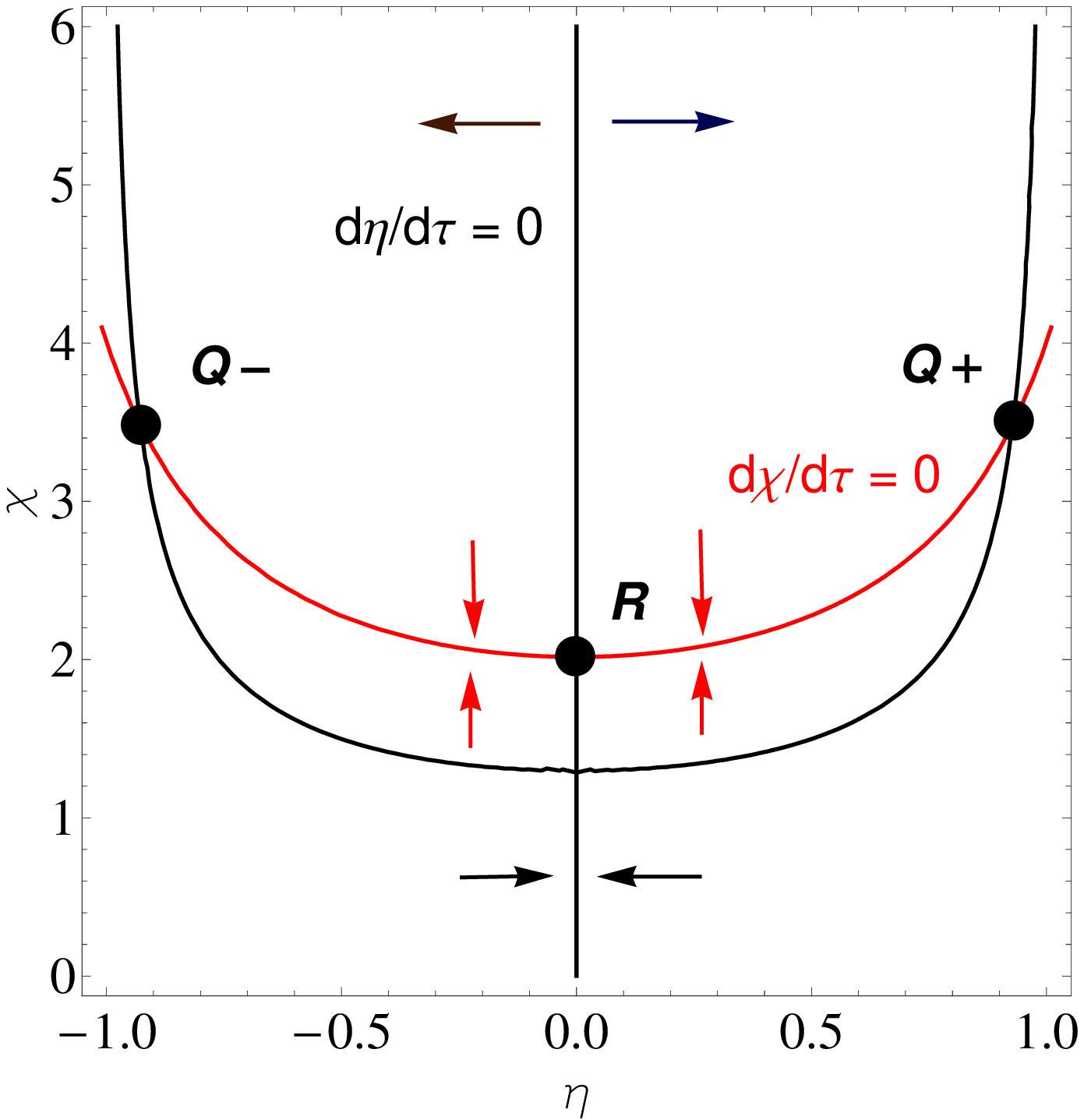}
\end{tabular}
\caption{\label{FullyOpenClines} Nullclines for the open
  system~(\ref{chidotopen},\ref{eeopen}). These curves correspond to
  $u = 0.3$ implying $g_{crit} = 0.59$. The $d \eta/d\tau= 0$ and
  $d\chi/d\tau = 0$ nullclines are plotted in black and red,
  respectively. The black (red) arrows indicate the regions of phase
  space where $\eta$ ($\chi$) increases or decreases. Left: $g = 0.79
  > g_{crit}$. The nullclines intersect in only the one point $R$
  representing the asymptotic stable racemic state. Right: $g = 0.29 <
  g_{crit}$. In this case there are three intersections, $Q_{\pm}$ and
  $R$ representing the $Z_2$ equivalent stable chiral states and the
  unstable racemic state, respectively. From Eq.~(\ref{eeu}) the
  enantiomeric excess at $Q_{\pm}$ is $\eta = \pm 0.93$.}
\end{figure}

In the phase space of the dynamical system defined by
Eqs.~(\ref{chidotopen},\ref{eeopen}) there are curves with a special
significance. These are the nullclines defined by
\begin{eqnarray}\label{nullclines}
\frac{d \chi}{d\tau} &=&  0,\\
\frac{d \eta}{d\tau} &=& 0.
\end{eqnarray}
The intersections of these curves give the possible steady states (or
fixed points) of the system.  The condition $d\chi/d \tau=0$ leads to
two curves
\begin{equation}
\chi_{\pm}^{(1)} = \frac{1 \pm \sqrt{1 + 4u(g(1+\eta^2)+ 1 -
\eta^2)}}{g(1+\eta^2)+ 1 - \eta^2},
\end{equation}
while $d\eta /d\tau = 0$ implies the three curves
\begin{equation}
\eta = 0, \qquad \chi_{\pm}^{(2)}= \pm \sqrt{\frac{4u}{1-g
+(g-1)\eta^2}}.
\end{equation}
For $u>0$ the solutions denoted $\chi_{-}$ correspond to negative
total enantiomer concentrations so we discard them. The physically
acceptable nullclines are plotted in Fig.~\ref{FullyOpenClines}.
Which of the two different intersection configurations is obtained
depends only on the single parameter $g$.
We emphasize that despite the similar appearance, the nullcline graphs
should not be confused with the classic bifurcation diagrams that have
been discussed often in the past~\cite{MKG,Kondepudi,Avetisov}.

%

%------------------------------------------------------------
\subsection{\label{sec:openstab}Fixed Points and Stability}
%------------------------------------------------------------

The system has several steady states: besides an unphysical state that
we disregard, there is a $Z_2$ pair of chiral solutions
$Q_{\pm}$, and a racemic state $R$:
\begin{eqnarray}
%U &=& \Big(P = \frac{2 (g+1) u-\sqrt{4 (g+1) u+1}+1}{2 (g+1)^2
%\gamma
%},\chi = \frac{1-\sqrt{4 (g+1) u+1}}{g+1},y= 0\Big),\\
R &=& \Big(P= \frac{2 (g+1) u+\sqrt{4 (g+1) u+1}+1}{2 (g+1)^2
\gamma },\chi = \frac{1+\sqrt{4 (g+1) u+1}}{g+1},y= 0\Big),\\
Q_{\pm} &=& \Big( P= \frac{u}{\gamma -g \gamma },\chi =\frac{1}{g},y= \pm
\frac{\sqrt{((g-1)/g^2)+4 u}}{\sqrt{g-1}}
\Big).
\end{eqnarray}
The associated eigenvalues are given by~\cite{RH}:
\begin{eqnarray}
%\lambda_{1,2,3}(U) &=& \Big(\sqrt{4 (g+1) u+1},\frac{1 + g\sqrt{4
%(g+1)
%u+1}}{g+1},-\gamma \Big),\\
\lambda_{1,2,3}(R) &=&\Big(-\sqrt{4 (g+1) u+1},\frac{1-g
\sqrt{4 (g+1) u+1}}{g+1},-\gamma \Big),\\
\lambda_{1,2,3}(Q_{\pm}) &=& \Big(\frac{-\sqrt{16 g^3 u+4 g^2-4
g+1}-1}{2 g},\frac{\sqrt{16 g^3 u+4 g^2-4 g+1}-1}{2 g},-\gamma
\Big).
\end{eqnarray}
Note that $\lambda_1(R) <0$ and $\lambda_1(Q) <0$ are always
negative whereas $\lambda_2(R) > 0$ and $\lambda_2(Q) < 0$ for $g <
g_{crit}$, otherwise $\lambda_2(R) < 0$ and $\lambda_2(Q) > 0$ for
$g > g_{crit}$, where $g_{crit} = (\sqrt{1+16u}-1)/8u$ is the
critical value for this parameter. Note that $g_{crit}(u) \leq 1$
for all $u \geq 0$. For small $u$ we can write $g_{crit} = 1 - 4u$;
while for large $u$, $g_{crit} \rightarrow 1/2u^{1/2}$. Thus the
direct chiral monomer production step ($\propto k_{1}$
in~(\ref{decay})) tends to racemize the system leading to final
$\eta$ values strictly less than unity:
\begin{equation}\label{eeu}
\eta = \pm\sqrt{1-\frac{4ug^2}{1-g}} \,\, ,
\end{equation}
which holds when $g < g_{crit}$. The chiral monomer production step
thus reduces the range of $g$ for which stable mirror symmetry
breaking can occur, and the chiral solutions are no longer $100\%$
chiral.

\begin{figure}[h]
\begin{tabular}{cc}
\includegraphics[width=0.45\textwidth]{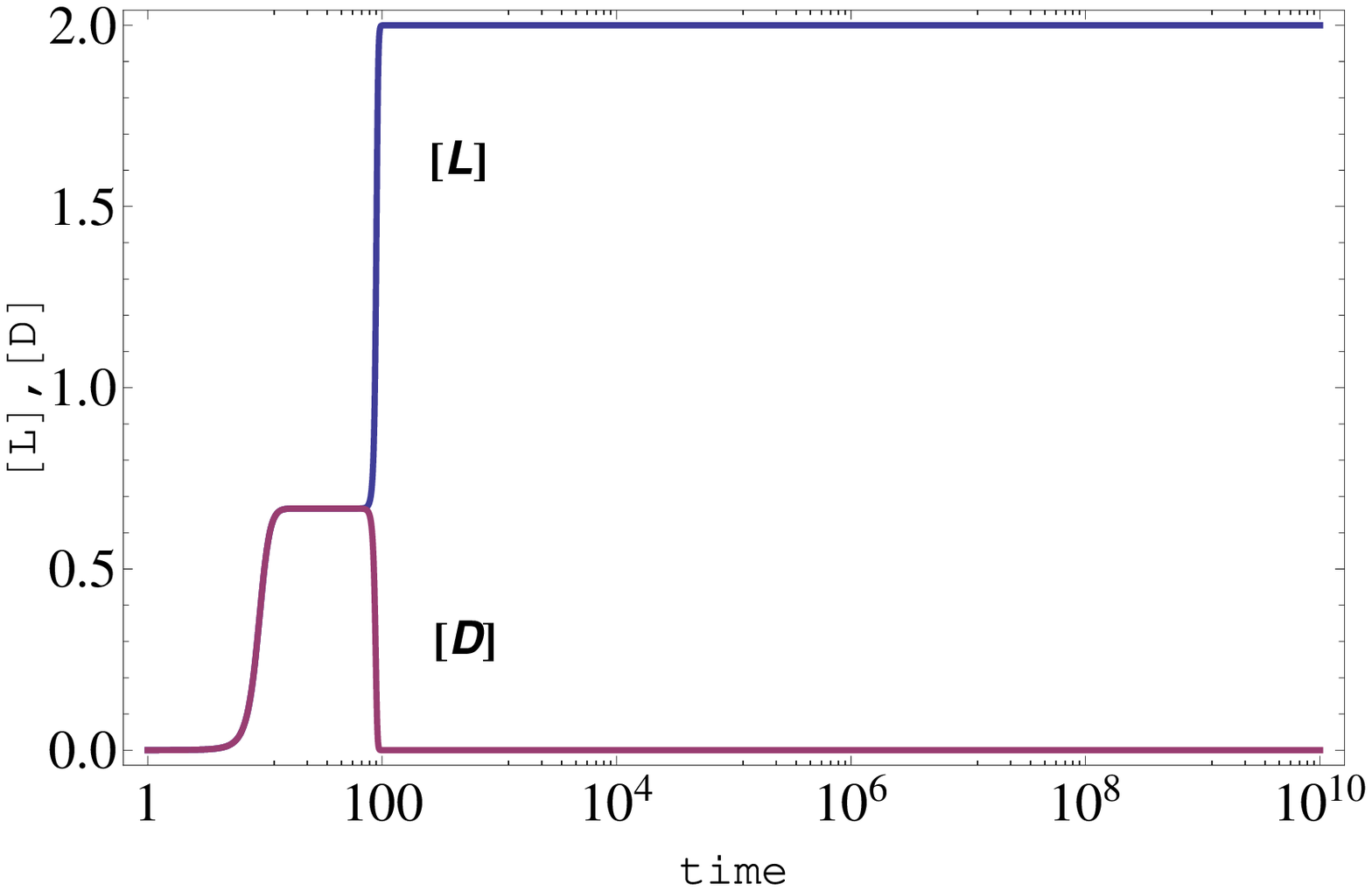}&
\includegraphics[width=0.45\textwidth]{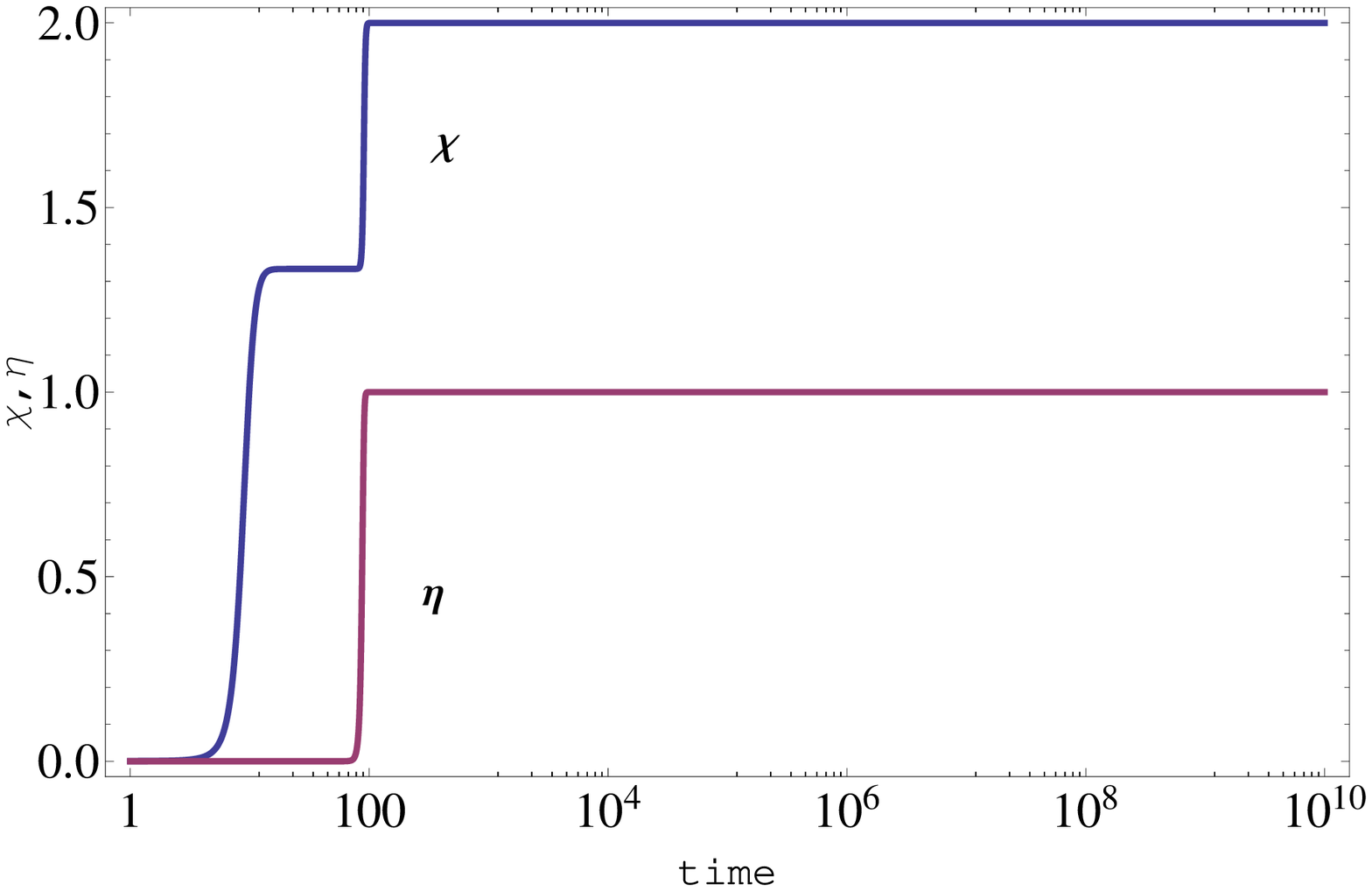}
\end{tabular}
\caption{\label{symbreaking} Chiral symmetry breaking in the open
  system~(\ref{Lopen}-\ref{LDopen}).  Temporal behavior (logarithmic
  scale) of the individual enantiomer concentrations $[L]$ and $[D]$
  (left), and the chiral polarization $\eta$ and total enantiomer
  concentration $\chi$ (right).  Initial concentrations: $[L]_0 = (1
  \times 10^{-6} + 1 \times 10^{-15})M$, $[D]_0 = 1 \times 10^{-6}
  M$ ($\eta_0 = 5 \times 10^{-8}\%$) and $[A]=1 M$. Rate constants:
  $k_1=10^{-4}s^{-1}$, $k_{-1}=10^{-6}s^{-1}$, $k_2 = 1s^{-1}M^{-1}$,
  $k_{-2}=0.5 s^{-1}M^{-1}$ and $k_{5}=1 s^{-1}M^{-1}$. These
  values correspond to $g=0.5$ and $u=10^{-4}$. In figures
  of simulations, we always display original variables $[L],[D],[LD]$,
  etc. as function of time $t$.}
\end{figure}
Figure~\ref{symbreaking} shows the temporal evolution of the L and D
chiral monomers starting from an extremely dilute total enantiomer
concentration and the very small statistical chiral deviations from
the ideal racemic composition. The right hand side of this figure
shows the evolution in terms of the quantities $\chi$ and $\eta$.
Note the mirror symmetry breaking signalled by  $\eta$.
%Compare this graph with the corresponding
%phase-space trajectory in Figure~\ref{ParamPlot}.
Are there chiral excursions found in the open system model? A chiral
excursion holds when the enantiomeric excess $\eta$ departs from a
small initial value, evolves to some maximum absolute value and then
decays to the final value of zero. To ensure a final racemic state
we must set $g>g_{crit}$ but then we find no numerical evidence for
such temporary chiral excursions. This can be understood
qualitatively from inspection of left hand side of
Fig.~\ref{FullyOpenClines}. The initial conditions (dilute chiral
monomer concentration and statistical chiral fluctuation)
corresponds to a initial point located at tiny values of $\chi$ and
close to the vertical nullcline, well below the point labeled as
$R$. The system is attracted to the black curve and moves up the
curve to $R$. In this situation, it is impossible for the chiral
excess to increase, not even temporarily. Notice the time scales for
$\chi$ and $\eta$ are of the same order. On the other hand, if $g <
g_{crit}$, then we have the situation depicted on the right hand
side of the figure. Here the same initial point moves towards the
vertical nullcline and up towards $R$, but once past the locally
horizontal black curve, is attracted to one of the two chiral fixed
points where it stays forever, provided the system is maintained out
of equilibrium. The chiral symmetry is permanently broken, and there
is no excursion such as we have defined it.

%------------------------------------------------------------
\section{\label{sec:semiopen} Semi-open system}
%------------------------------------------------------------

%------------------------------------------------------------
\subsection{\label{sec:semiopen1} Rate equations}
%------------------------------------------------------------

To elucidate the temporal evolution of $\chi$ and $\eta$ for a more
general setting, we do not remove the heterodimer from the system
now allow the back reaction to chiral monomers, and we keep [A]
constant. There is an implicit inflow as a consequence of constant
$[A]$, but no outflow, so we denote this case "semi-open". While
there is no mass balance the system can still exhibit temporary
SMSB.

The corresponding rate equations are
\begin{eqnarray}\label{Lsemi}
\frac{d}{d t}[L] &=& k_1[A] + (k_2[A]-k_{-1})[L]-k_{-2}[L]^2 -
k_5[L][D] + k_{-5}[LD], \\
\label{Dsemi} \frac{d}{d t}[D] &=& k_1[A] + (k_2[A]-k_{-1})[D]
-k_{-2}[D]^2 - k_5[D][L] + k_{-5}[LD], \\
%\label{Asemi} \frac{d}{dt}[A] &=&
%-2k_1[A]-(k_2[A] - k_{-1})([L]+[D])+ k_{-2}([L]^2 + [D]^2), \\
\label{LDsemi} \frac{d}{d t}[LD] &=& k_5[L][D] - k_{-5}[LD].
\end{eqnarray}
After performing the same rescaling as in the open case, we arrive at
\begin{eqnarray}\label{chidot2}
\frac{d \chi}{d \tau} &=& 2u + \chi
-\frac{1}{2}(g+1)\chi^2 -\frac{1}{2}(g-1)\chi^2 \eta^2 + 2 r P,\\
\label{ee2} \frac{d \eta}{d \tau} &=& \eta (1 - g \chi)
-\frac{\eta}{\chi}\left(\frac{d \chi}{d \tau}\right),\\ \label{Pdot2}
\frac{d P}{d \tau} &=&\frac{1}{4}\chi^2 (1 - \eta^2) - r P.
\end{eqnarray}
The dimensionless parameters appearing here are:
\begin{equation}\label{paramsemiopen}
u = \frac{k_1 k_{5}[A]}{(k_2[A] - k_{-1})^2},\,\, g =\frac{k_{-2}}{k_5},
\,\, r = \frac{k_{-5}}{(k_2[A] - k_{-1})}.
\end{equation}
The system is described by three equations
Eqs.~(\ref{chidot2}-\ref{Pdot2}) which do not decouple.

%------------------------------------------------------------
\subsection{\label{sec:semiopennull} Phase plane and linear stability analysis}
%------------------------------------------------------------

In order to obtain an \textit{approximate} two-dimensional phase
plane representation of the system, we will invoke the dynamic
steady state approximation for the heterodimer. Such approximations
are usually justified when there exists a clear separation of time
scales in the problem, thus allowing one to identify rapidly and
slowly changing concentrations~\cite{Scott}. Here however, no such
time scales are evident, all concentration variables evolve on a
similar time scale.  Nevertheless, we will see a posteriori that
this approximation can be good over a wide range of time scales. We
therefore assume that the heterodimer is in a dynamic steady state
$P_{stat}$ relative to the chiral monomer concentrations and chiral
polarization:
\begin{equation}\label{Psteady2}
P_{stat} \approx \frac{\chi^2}{4r}(1 - \eta^2).
\end{equation}
Substituting this $P_{stat}$ into Eqs.~(\ref{chidot2},\ref{ee2}) leads
to the differential equations
%%
%\begin{eqnarray}\label{f}
%\frac{d \chi}{d \tau} &=& f(\chi, \eta),\\ \label{g}
%\frac{d \eta}{d
%\tau} &=& g(\chi,\eta),
%\end{eqnarray}
%%
%where
%
\begin{eqnarray}\label{chisemi}
\frac{d \chi}{d \tau} &=& 2u + \chi - \frac{g}{2}\chi^2(1 +
\eta^2),\\
\frac{d \eta}{d \tau} %&=& \eta(1 - g\chi)
%-\frac{\eta}{\chi}f(\chi,\eta)\\
&=& \eta \Big(-g\chi -\frac{2u}{\chi} + \frac{g}{2}\chi(1 + \eta^2)
\Big).\label{etasemi}
\end{eqnarray}

\begin{figure}[h]
\begin{tabular}{cc}
\includegraphics[width=0.40\textwidth]{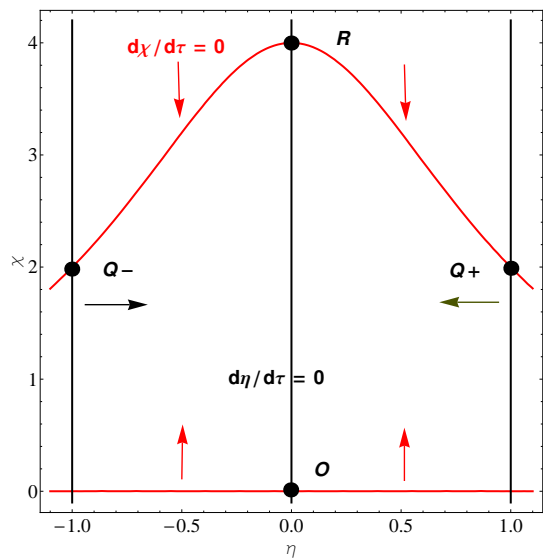}&
\end{tabular}
\caption{\label{semiopencline} %Left:
  Nullclines for the semi-open case~(\ref{chisemi},\ref{etasemi}) in
  the steady state approximation for $P$.  The
  nullclines~(\ref{fsclines},\ref{gsclines}) are plotted in red and
  black, respectively. Red and black arrows indicate the phase-space
  regions of increasing or decreasing $\chi$ and $\eta$. The four
  stationary solutions $O,R,Q_{\pm}$ are indicated by the black dots.
  These curves are illustrated for $u=0$ and $g=0.5$.}
\end{figure}

As above, we study the phase space of the two-dimensional system by
means of the nullclines. These are plotted in Fig.
\ref{semiopencline}. The condition $d\chi /d\tau=0$ implies two
curves
\begin{equation}\label{fsclines}
\chi_{\pm} = \frac{1 \pm \sqrt{1 + 4gu(1+\eta^2)}}{g(1+\eta^2)},
\end{equation}
whereas the condition $d\eta /d\tau = 0$ implies the three curves
\begin{equation}\label{gsclines}
\eta = 0, \qquad \eta_{\pm} = \pm \sqrt{1 + \frac{4u}{g \chi^2}}.
\end{equation}
%

%------------------------------------------------------------
\subsection{\label{sec:semiopenstab}Fixed Points and Stability}
%------------------------------------------------------------

We solve
Eqs.~(\ref{chidot2}-\ref{Pdot2}) looking for
steady states. To keep the algebra manageable, we also
set $u=0$ as in~\cite{RH}. There are four solutions, namely,
the empty $O$ solution, the racemic $R$ or the two mirror-symmetric
chiral $Q_{\pm}$ solutions:
\begin{eqnarray} %\begin{array}{lll}
O & = & (P=0, \chi = 0 , y = 0), \\
R & = & \left( P= \frac{1}{g^2 r},\chi = \frac{2}{g} , y = 0 \right), \\
Q_{\pm} & = & \left( P = 0, \chi = \frac{1}{g}  , y = \pm \frac{1}{g} \right).
%\end{array}
\end{eqnarray}
Note that the final heterodimer concentration $P$ is \textit{zero}
in the chiral states $Q_{\pm}$. Note that the steady state
approximation Eq.~(\ref{Psteady2}) implies the same result since
$|\eta| = 1$.
In order to study the linear stability of the four possible
homogeneous solutions $O, R$ and $Q_{\pm}$, we calculate the
eigenvalues of the $3\times 3$ Jacobian array $M_{open}$ derived
in~\cite{RH} after deleting the 3rd and 4th rows and columns. The
eigenvalues corresponding to these solutions are given by
\begin{eqnarray}
  \lambda_{1,2,3}(O) &=& (1,1,-r),  \\
  \lambda_{1,2,3}(R) &=&
  \Big(-1,\,-\frac{2+g(1+r)+\sqrt{4+g^2(-1+r)^2+4g(1+r)}}{2g},
   \nonumber \\
  && \frac{-2-g(1+r)+\sqrt{4+g^2(-1+r)^2+4g(1+r)}}{2g} \Big).  \\
  \lambda_{1,2,3}(Q_{\pm}) &=& \Big(-1,\,-\frac{1+g(-1+r)+\sqrt{1+2g(-1+r)+g^2(1+r)^2}}{2g}, \nonumber \\
   && \frac{-1+g(1-r)+\sqrt{(1+g(-1+r))^2+4g^2r}}{2g}\Big).
\end{eqnarray}
As $\lambda_{1,2}(O) >0$, the empty state is always unstable. An
inequality analysis shows that both $\lambda_2(R) <0$ and
$\lambda_3(R) <0$ for all $r>0$ and $g>0$. Since $\lambda_1(R) = -1$
this demonstrates that the racemic state $R$ is always stable. As an
independent check, we also verify that $\lambda_3(Q) > 0$ is
positive for all $r>0$ and $g>0$, so the chiral solutions $Q_{\pm}$
are always unstable. The final outcome will always the racemic
state. There is no stable mirror symmetry broken solution when the
heterodimer dissociates back into the chiral monomers. Nevertheless,
the system can have temporary chiral excursions.

\begin{figure}[h]
\begin{tabular}{cc}
\includegraphics[width=0.45\textwidth]{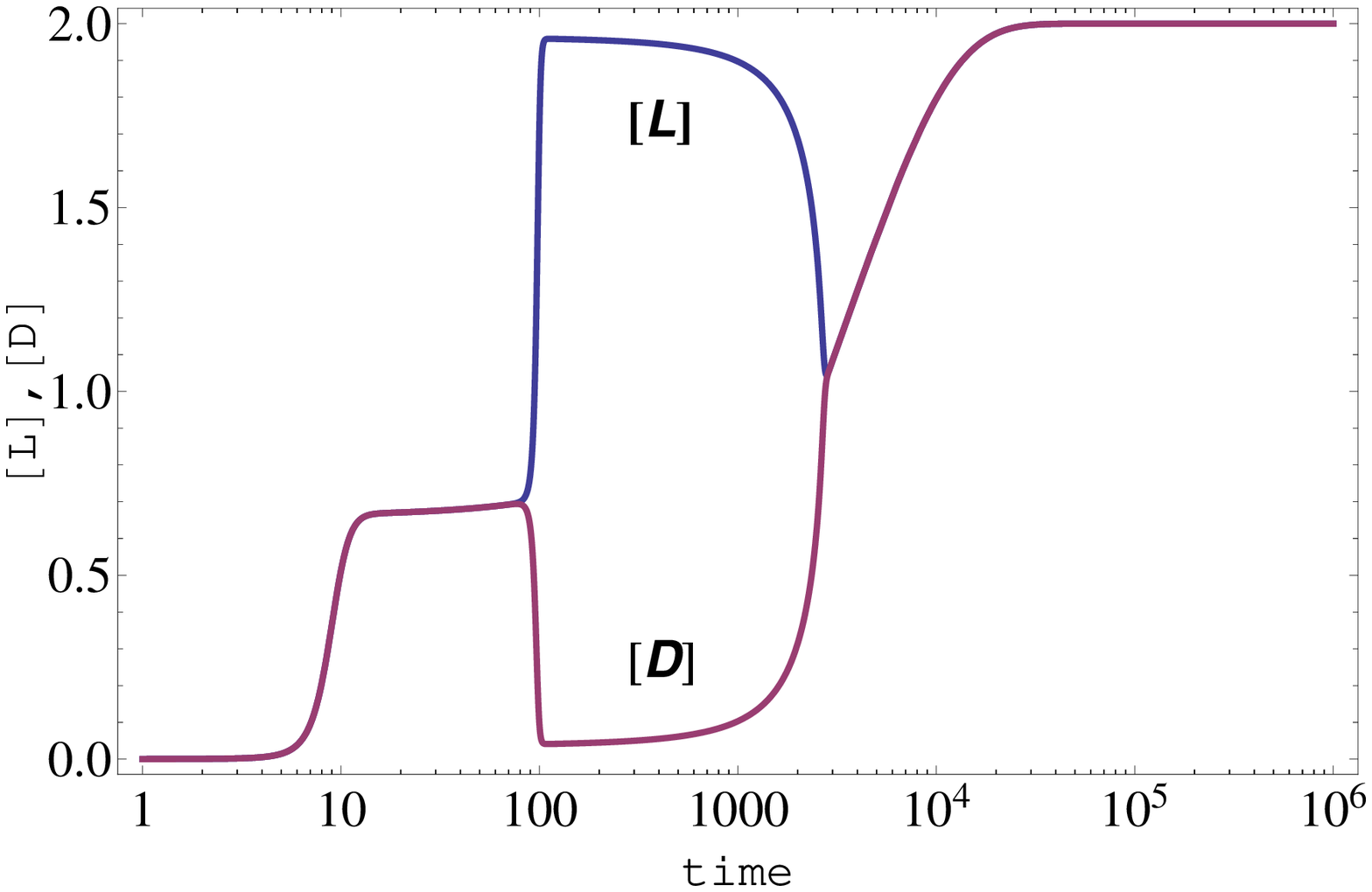}&
\includegraphics[width=0.45\textwidth]{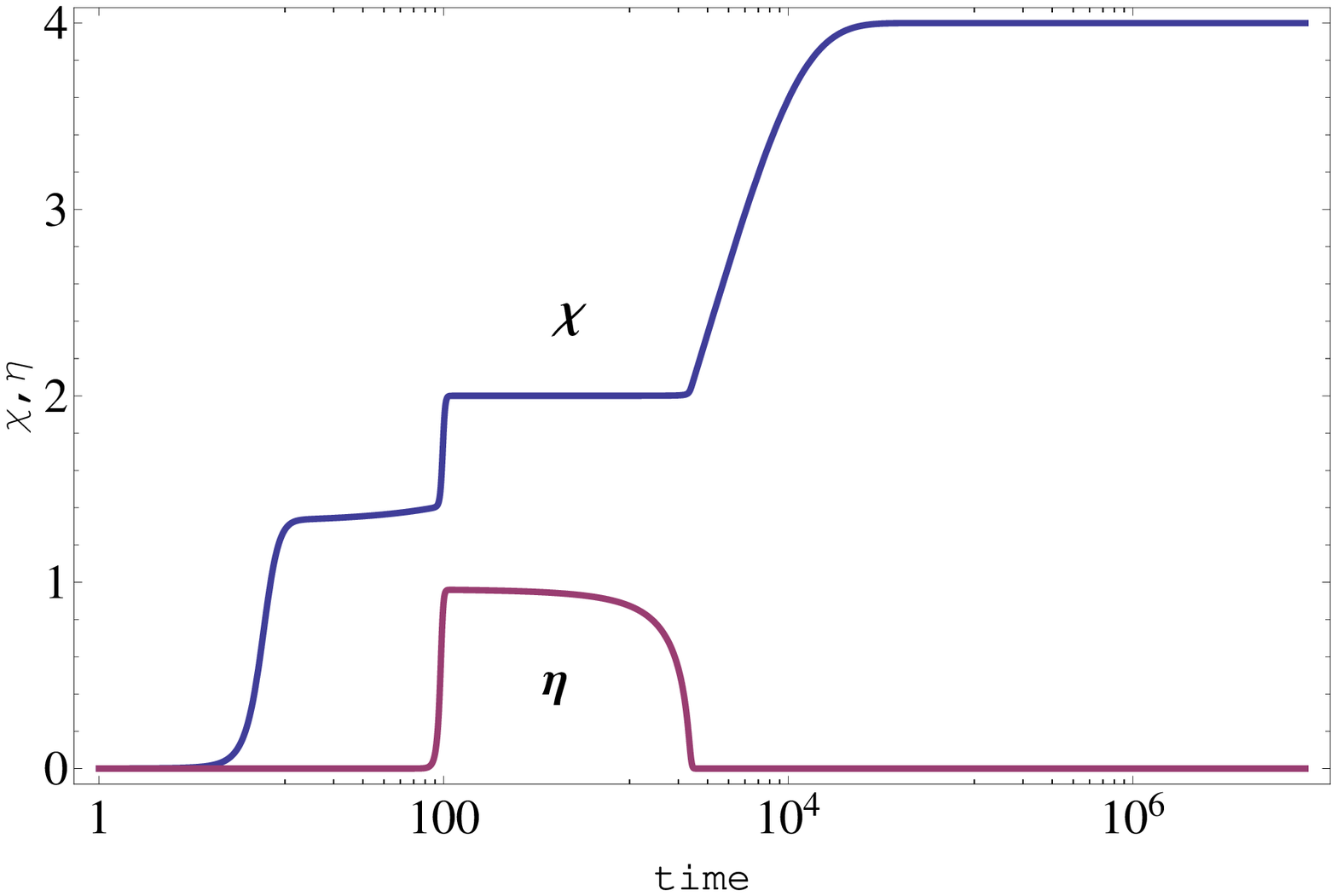}%\newline
%\hspace{0.25\textwidth}
%\includegraphics[width=0.50\textwidth]{SemiOpenHeterodimers.eps}
\end{tabular}
\caption{\label{excursionsO} Temporary chiral symmetry breaking and
  chiral excursions in the
  semi-open system~(\ref{Lsemi}-\ref{LDsemi}).
Temporal behavior (logarithmic scale) of the individual enantiomer
concentrations $[L]$ and $[D]$ (left) and the chiral polarization
$\eta$ and total enantiomer concentration $\chi$ (right). Initial
concentrations: $[L]_0 = (1 \times 10^{-7} + 1 \times 10^{-15})M$,
$[D]_0 = 1 \times 10^{-7} M$ ($\eta_0 = 5 \times 10^{-8}\%$) and
$[A]=1 M$. Rate constants: $k_{1}=10^{-4}s^{-1}$,
$k_{-1}=10^{-6}s^{-1}$, $k_2 = 1s^{-1}M^{-1}$, $k_{-2}=0.5
s^{-1}M^{-1}$, $k_{5}=1 s^{-1}M^{-1}$ and $k_{-5}=10^{-3}s^{-1}$.
These rate constants imply $g=0.5$ and $u=10^{-4}$.}
\end{figure}
\begin{figure}[h]
\includegraphics[width=0.50\textwidth]{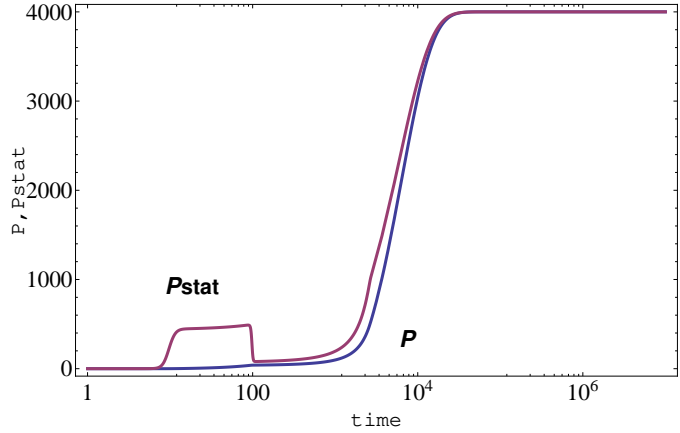}
\caption{\label{PstatO} Semi-open system: Comparison of direct
  numerical solution $P = [\widetilde{LD}]$ and the steady state approximation $P_{stat}$
  in Eq.~(\ref{Psteady2}) for the heterodimer concentration (after
  transforming from $[L],[D],[LD]$ to $\chi,\eta,P$).}
\end{figure}

In Fig.~\ref{excursionsO} we plot the temporal evolution of the $L$
and $D$ chiral monomers starting from an extremely dilute total
enantiomer concentration and the very small statistical chiral
deviations from the ideal racemic composition. The right hand side
of this figure shows the evolution in terms of the quantities $\chi$
and $\eta$. Note the chiral excursion in $\eta$ for the time
interval between $t\simeq 100s$ and $t\simeq 1000s$. Finally, we
compare the heterodimer concentration from direct numerical
simulation with the steady state approximation in Fig.~\ref{PstatO}.
To do so we simulate the original concentration variables in
Eqs.(\ref{Lsemi}-\ref{LDsemi}) and then transform results in terms
of $\chi$, $\eta$ and $P$.  While $P_{stat}$ appears to overestimate
$P = [\widetilde{LD}]$, it provides a reasonably good approximation
to the actual heterodimer concentration $P$ right after chiral
symmetry is broken, at about $t \simeq 100 s$ and coincides
perfectly after chiral symmetry is recovered, and when $\chi$
reaches its asymptotic value, after approximately $t \simeq 10^4 s$.

%------------------------------------------------------------
\section{\label{sec:closedsystems}Closed system}
%------------------------------------------------------------

%------------------------------------------------------------
\subsection{\label{sec:closedsystems1}Rate equations}
%------------------------------------------------------------

The rate equations directly follow from~(\ref{decay}-\ref{heterodimer}):
\begin{eqnarray}\label{L}
\frac{d}{d t}[L] &=& k_1[A] + (k_2[A]-k_{-1})[L]-k_{-2}[L]^2 -
k_5[L][D] + k_{-5}[LD], \\
\label{D} \frac{d}{d t}[D] &=& k_1[A] + (k_2[A]-k_{-1})[D]
-k_{-2}[D]^2 - k_5[D][L] + k_{-5}[LD], \\
\label{A} \frac{d}{dt}[A] &=&
-2k_1[A]-(k_2[A] - k_{-1})([L]+[D])+ k_{-2}([L]^2 + [D]^2), \\
\label{LD} \frac{d}{d t}[LD] &=& k_5[L][D] - k_{-5}[LD].
\end{eqnarray}
There is no flow of material into or out of the system. Since $[A]$
is not constant in this situation, we cannot use it to rescale the
time or the concentrations. Instead, we take $\tau = k_{1} t$ for the
time parameter and $[\tilde L] = (k_{5}/k_{1}) [L]$, etc. for the
dimensionless concentrations. This allows us to express the rate
equations in the following dimensionless
form:
\begin{eqnarray}\label{eqntilL}
\frac{d}{d \tau}[\tilde L] &=& [\tilde A] - u[\tilde L] + h[\tilde
A][\tilde L] - g[\tilde L]^2 -[\tilde L][\tilde D] + \rho[\widetilde{LD}], \\
\label{eqntilD} \frac{d}{d \tau}[\tilde D] &=& [\tilde A] - u[\tilde
D] + h[\tilde A][\tilde D] - g[\tilde D]^2 -[\tilde D][\tilde L] +
\rho[\widetilde{LD}],\\
\frac{d}{d \tau}[\widetilde{LD}] &=& [\tilde L][\tilde D] -
\rho[\widetilde{LD}].
\end{eqnarray}
These are subject to the constraint $[\tilde A] = [\tilde C] -
([\tilde L]+[\tilde D])-2[\widetilde{LD}]$, where $[\tilde C]  =
(k_{5}/k_{1}) [C]$ and $[C]$ is the total initial
concentration, being constant in time. The four parameters appearing here are
\begin{eqnarray}\label{paramsclosed}
u = \frac{k_{-1}}{k_{1}},\, g =\frac{k_{-2}}{k_{5}}, \, h =
\frac{k_{2}}{k_{5}}, \, \rho = \frac{k_{-5}}{k_{1}}.
\end{eqnarray}
Changing variables as before to $\chi, \eta, P$, we arrive at
\begin{eqnarray}\label{chidot}
\frac{d \chi}{d \tau} &=& 2[\tilde A] + (h[\tilde A] - u)\chi
-\frac{1}{2}(g+1)\chi^2 -\frac{1}{2}(g-1)\chi^2 \eta^2 + 2\rho P,\\
\label{ee} \frac{d \eta}{d \tau} &=& \eta (h[\tilde A] - g \chi -u)
-\frac{\eta}{\chi}\left(\frac{d \chi}{d \tau}\right),\\ \label{Pdot}
\frac{d P}{d \tau} &=&\frac{1}{4}\chi^2 (1 - \eta^2) - \rho P.
\end{eqnarray}
In these variables, the constant mass constraint reads $[\tilde A] =
[\tilde C] - \chi - 2P$.

%------------------------------------------------------------
\subsection{\label{sec:closednull}Phase plane and linear stability analysis}
%------------------------------------------------------------

As in the semi-open case, to obtain an approximate two-dimensional
phase plane portrait, we assume that the heterodimer is in an
approximate steady state $P_{stat}$ and solve Eq.~(\ref{Pdot}) for
\begin{equation}\label{Psteadyclosed}
P_{stat} \approx \frac{\chi^2}{4 \rho}(1 - \eta^2).
\end{equation}
Substituting this back into Eqs.~(\ref{chidot},\ref{ee}), we obtain
%Eqs.~(\ref{f},\ref{g}), where
%
\begin{eqnarray}\label{chiclosed}
\frac{d \chi}{d \tau} &=& (2+h\chi)[\tilde A]- u\chi - \frac{g}{2}\chi^2(1 +
\eta^2),\\
\frac{d \eta}{d \tau} %&=& \eta (h[\tilde A] - g\chi -u)
%-\frac{\eta}{\chi}f(\chi,\eta) \nonumber \\
&=& \eta \Big(-g\chi - \frac{2 [\tilde A]}{\chi} + \frac{g}{2}\chi(1 +
\eta^2)\Big),\label{etaclosed}
\end{eqnarray}
where
\begin{eqnarray}
[\tilde A]  &=& [\tilde C] - \chi -\frac{\chi^2}{2\rho}(1 - \eta^2).
\end{eqnarray}
\begin{figure}[h]
\includegraphics[width=0.50\textwidth]{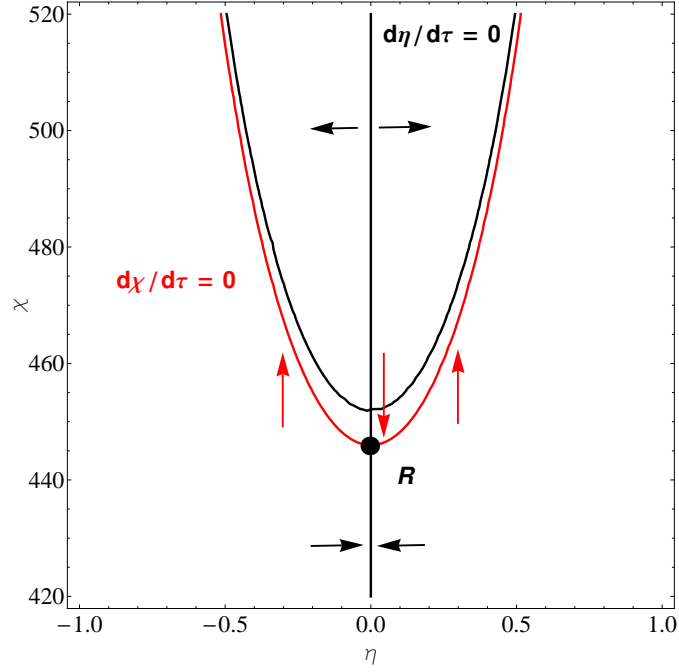}
\caption{\label{closedcline} Nullclines for the closed
  system~(\ref{chiclosed},\ref{etaclosed}) in the steady state
  approximation for $P$.  The $\eta$ and $\chi$
  nullclines are plotted as the black and red curve, respectively.
  These nullclines intersect at the one point (indicated
  with a dot) which corresponds to the stable racemic solution $R$
  with $\eta = 0$ and $\chi > 0$. The flow directions are indicated
  with arrows. This diagram corresponds to the parameter values
  $g=0.05$, $u=0.01$, $h=0.1$ and $\rho=1.0$ and $[\tilde C] = 10^5$.}
\end{figure}

The nullcline condition $d\chi /d \tau = 0$ leads to an unwieldy
cubic equation in $\chi$. More importantly, the nullcline is an
\textit{even} function of $\eta$, reflecting the underlying $Z_2$
mirror symmetry.  The other condition $d\eta /d \tau =0$ is
straightforward to solve analytically and leads -- after discarding
the unphysical solution corresponding to negative total enantiomer
concentrations -- to two curves
\begin{equation}\label{gnullexplicit}
\eta = 0 , \qquad \chi = \frac{-1 + \sqrt{1 + 2{[\tilde
C]}(g(1+\eta^2)/2 + (1-\eta^2)/\rho - g)}}{g(1+\eta^2)/2 +
(1-\eta^2)/\rho - g}.
\end{equation}
The curve $\chi(\eta)$ is an even function of $\eta$ and $\chi(\eta)
\rightarrow \infty$ for $\eta^2 \rightarrow 1$ with a minimum at $\eta
= 0$. Thus the nullcline has the form of a narrow fork
as depicted in Fig.~\ref{closedcline}.

\begin{figure}[h]
\begin{tabular}{cc}
\includegraphics[width=0.45\textwidth]{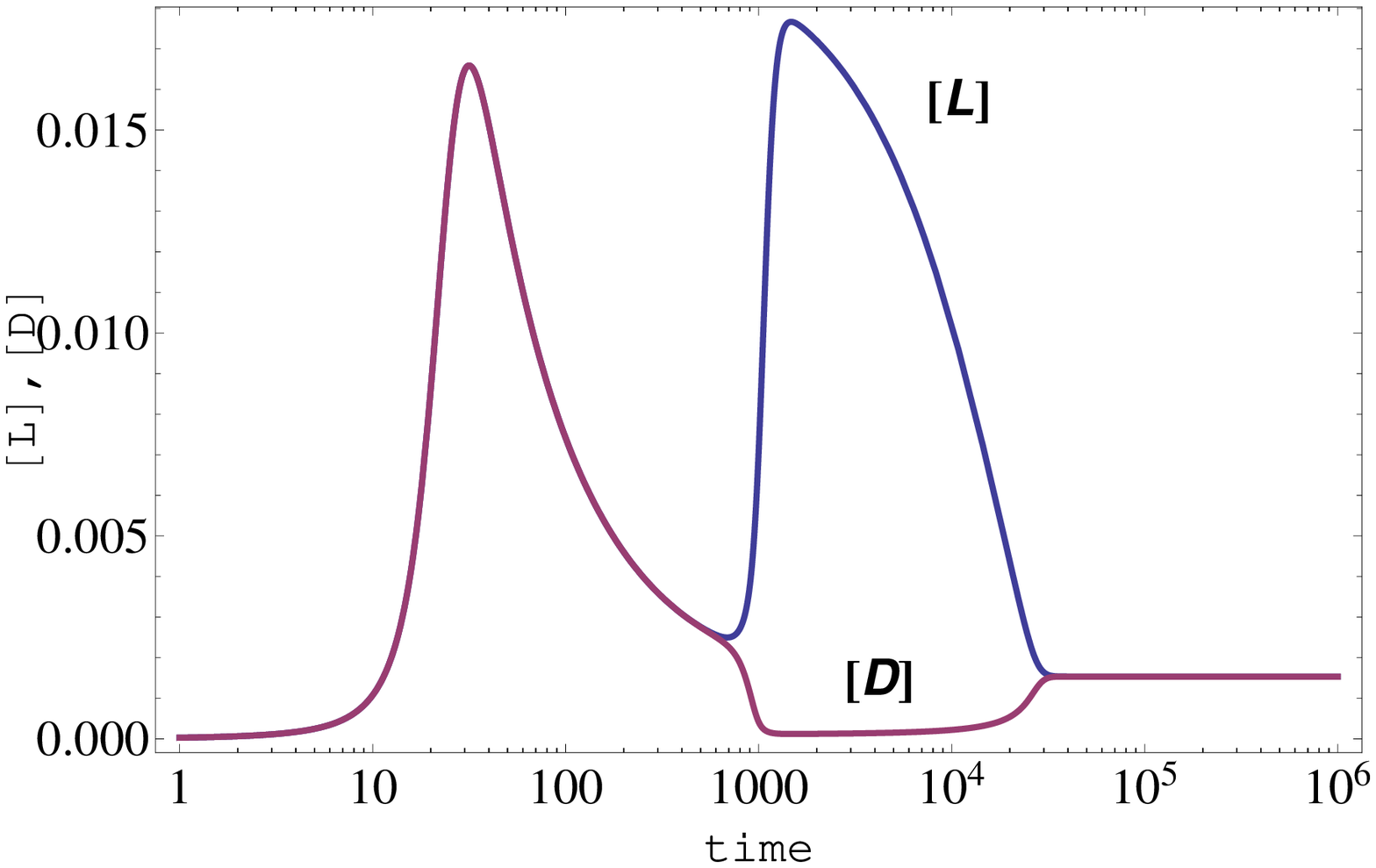}&
\includegraphics[width=0.45\textwidth]{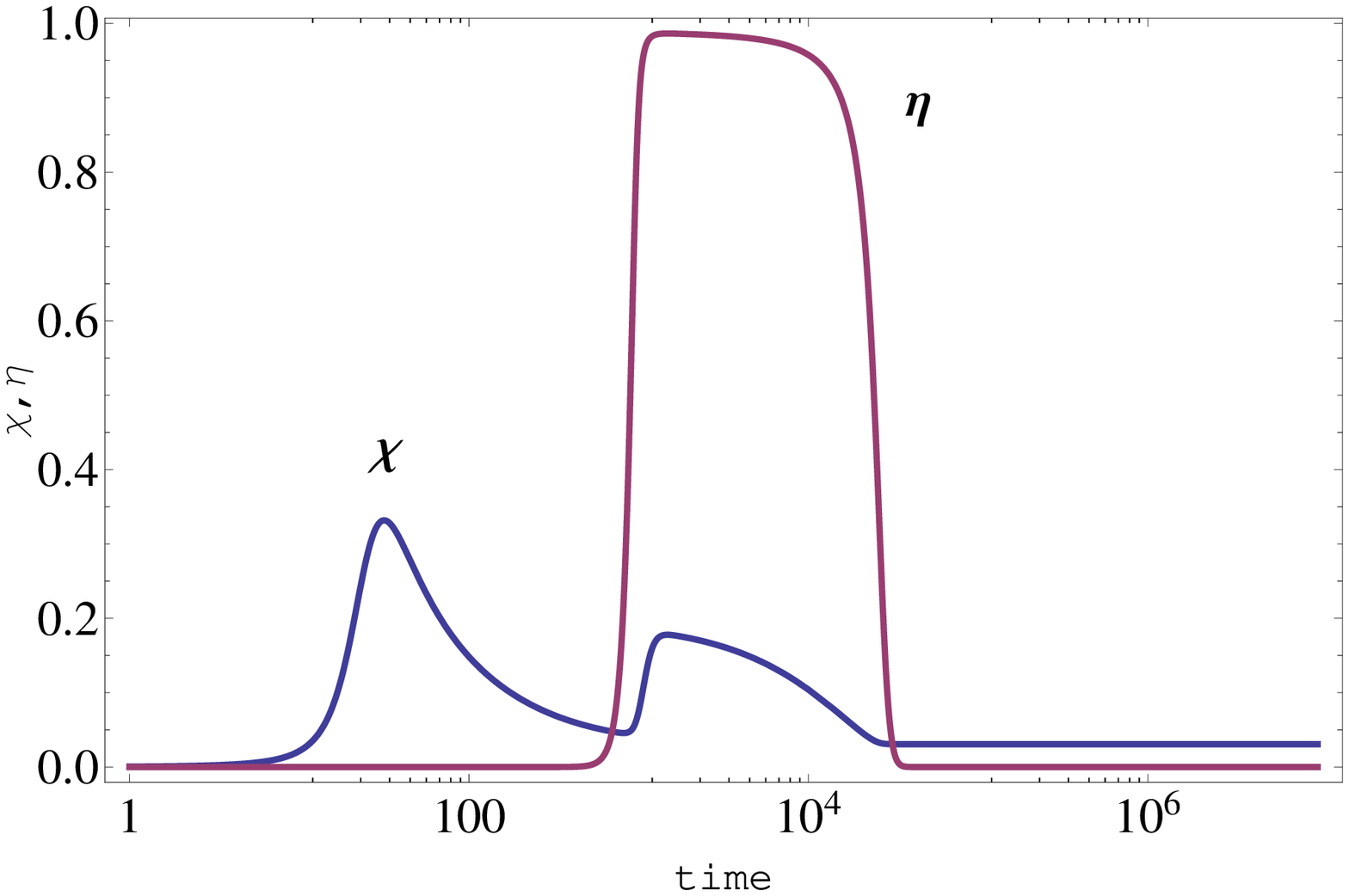}
\end{tabular}
\caption{\label{excursionclosed} Temporary chiral symmetry breaking
  and chiral excursions in
  the closed system~(\ref{L}-\ref{D}).
Temporal behavior (logarithmic scale) of the individual enantiomer
concentrations $[L]$ and $[D]$ (left) and the chiral polarization
$\eta$ and total enantiomer concentration $\chi$ (right). Initial
concentrations: $[L]_0 = (1 \times 10^{-7} + 1 \times 10^{-15})M$,
$[D]_0 = 1 \times 10^{-7} M$ ($\eta_0 = 5 \times 10^{-8}\%$) and
$[A]_0=1 M$. Rate constants: $k_{1}=10^{-4}s^{-1}$,
$k_{-1}=10^{-6}s^{-1}$, $k_2 = 1s^{-1}M^{-1}$,
$k_{-2}=0.5s^{-1}M^{-1}$, $k_{5}=10s^{-1}M^{-1}$ and
$k_{-5}=10^{-4}s^{-1}$.}
\end{figure}
\begin{figure}[h]
\includegraphics[width=0.50\textwidth]{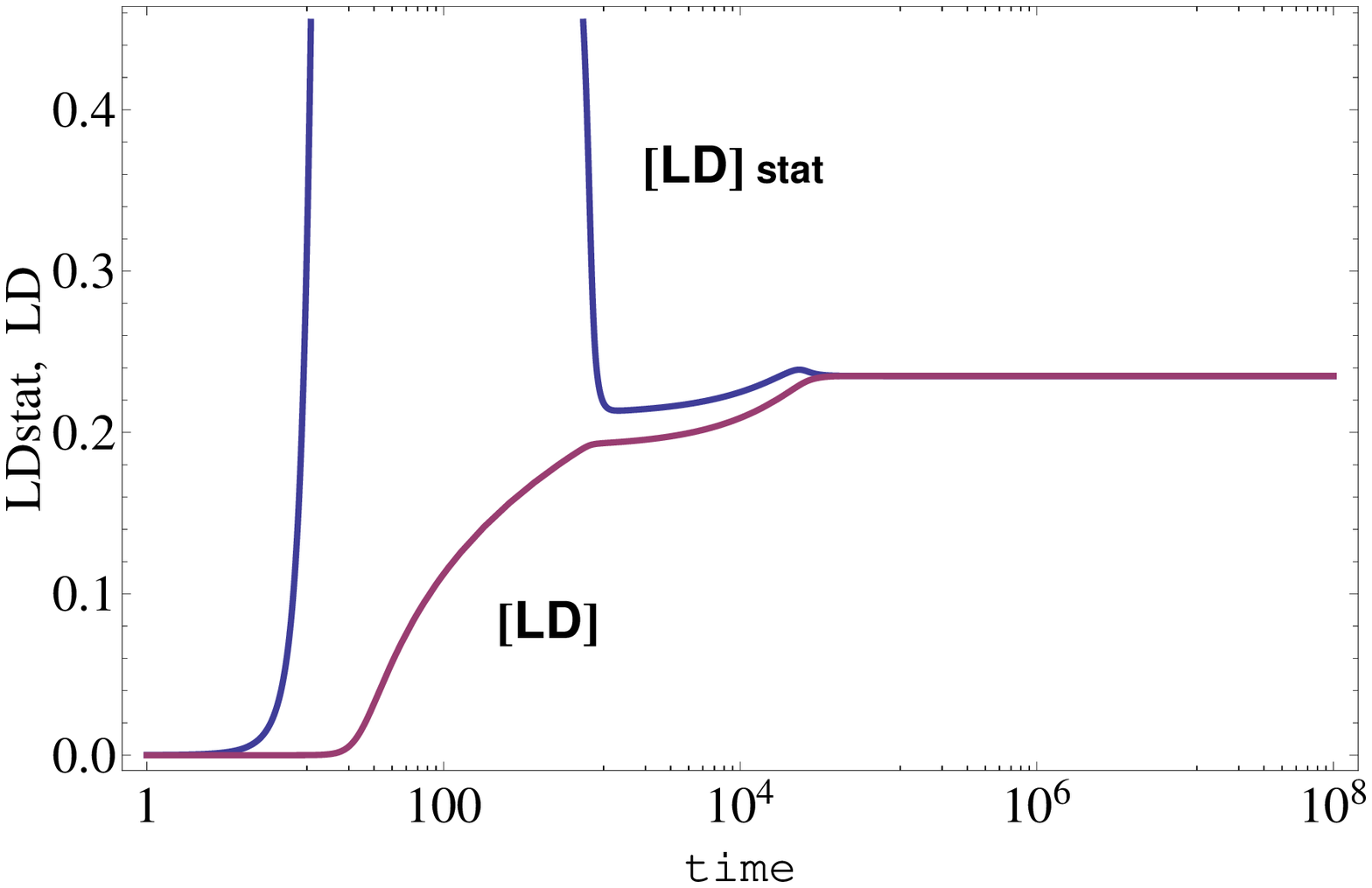}
\caption{\label{Pstat} Comparison of direct numerical solution
$[LD]$ and the steady state approximation $[LD]_{stat}=
(k_5/k_{-5})[L][D]$ for the heterodimer concentration. The steady
state approximation peaks to a maximum value of about 25 in this
example.}
\end{figure}
%

%------------------------------------------------------------
\subsection{\label{sec:closedstab}Fixed Points and Stability}
%------------------------------------------------------------

A linear stability analysis for the closed Frank model is given
in~\cite{RH}. That analysis was carried out in terms of $\chi,y $ and
$P$ and does not assume the stationary approximation for the
heterodimer. It turns out, even in a model as simple as this one, that
keeping $\rho > 0$ is analytically untractable, so we consider $\rho =
0$ in what follows.  Then the asymptotic stationary racemic $R$ and
chiral solutions $Q_{\pm}$ are given by
\begin{eqnarray}
R &=& \Big(P=\frac{[\tilde C]}{2}, \chi = 0, y=0\Big),\\
Q_{\pm} &=& \Big(P=\frac{[\tilde C] g + u}{2g},
\chi = -\frac{u}{g}, y = \pm \frac{u}{g}\Big).
\end{eqnarray}
The chiral solutions $Q_{\pm}$ are unphysical for all $u > 0$ since
they imply negative total enantiomer concentrations $\chi < 0$.
Thus, the only physically acceptable solution is the racemic one
$R$, and this is (at least marginally) stable, the corresponding
eigenvalue was calculated to be~\cite{RH}
\begin{equation}
\lambda_{1,2,3}(R) = (0,-2-u,-u).
\end{equation}
In the limit $\rho = 0$, the substrate is consumed and all the matter
ends up finally as pure heterodimer. Finally, note that $\lim_{u
  \rightarrow 0} Q_{\pm} = R$: the unphysical chiral solutions merge
to the racemic one when $k_{-1} = 0$. For reversible heterodimer,
the matter in the racemic state $\eta = 0$ is distributed between
the chiral monomers and the heterodimer: $P=([\tilde C]-\chi)/2$ and
$\chi
> 0$ in keeping with the law of mass action. The single intersection
$R$ displayed in Fig.~\ref{closedcline} indicates that the racemic
state is the only possible solution, in qualitative agreement with the
stability analysis.

Example of a chiral excursion in a closed system is provided in
Fig.~\ref{excursionclosed}. Once again, as for the semi-open
situation, the scheme is capable of amplifying a tiny initial chiral
excess to practically $100\%$, followed by final approach
to the racemic state. The steady state approximation for the heterodimer is
rather poor during the early stages (Fig.~\ref{Pstat}), but similar to
semi-open case, converges to the true heterodimer concentration after
symmetry breaking and restoration.

%------------------------------------------------------------
\section{\label{sec:disc} Discussion}
%------------------------------------------------------------

In this Letter, we investigated transient mirror symmetry breaking
in chiral systems, in particular in the Frank model in settings of
open, semi-open, and closed environments. Temporary chiral
excursions are observed for closed and semi-open systems and
explained through phase space analysis, stability analysis and
numerical simulations. Such chiral excursions may be experimentally
observed and could be mistaken for a transition to a chiral state.
They are in fact a long sought goal for the experimental chemist who
could actually fail to see them if not aware of their transitory
nature. In open systems by contrast, the racemic state is approached
monotonically. Therefore, it is important to understand the
processes and constraints responsible for these outcomes. This
Letter has focused on the effects that the in- and outflow of matter
has on these phenomena. The open nature of the Frank model can be
arranged \textit{experimentally} with an incoming flow of achiral
precursor A and an outflow of the product heterodimer LD, to
conserve mass balance. Mathematically, we can model the inflow by
assuming a constant concentration of $[A]$ and the outflow by a term
representing the rate at which LD leaves the system. In our open
model we assumed no dissociation of LD back into chiral monomers.
From the point of view of achieving permanent SMSB, there is then
actually no need to remove LD from the system, but we retain this
outflow since it is needed to ensure stationary fixed points for all
the chemical concentrations. For the semi-open case, LD is not
removed and we allow for its dissociation into chiral monomers.
There is no mass balance but temporary symmetry breaking can arise.
Finally, in the closed system there is neither inflow nor outflow,
total mass is conserved and temporary symmetry breaking can occur.

A recent kinetic analysis of the Frank model in closed systems applied
to the Soai reaction~\cite{Soai} indicates that in actual chemical
scenarios, reaction networks that exhibit SMSB are extremely sensitive
to chiral inductions due to the presence of inherent tiny initial
enantiomeric excesses~\cite{CHMR}. This amplification feature is also
operative in much more involved reaction networks such as chiral
polymerization~\cite{BH}. When the system is subject to a very small
perturbation about an extremely dilute racemic state, the initial
chiral fluctuation does not immediately decay, but becomes amplified
and drives the system along a long-lived chiral excursion in phase
space before final and inevitable approach to the stable racemic
solution.  Mauksch and Tsogoeva have also previously indicated that
chirality could appear as the result of a temporary asymmetric
amplification~\cite{MS1,MS2}.

Excursions in phase space as studied in this work are superficially
reminiscent of excitable systems as studied in dynamical
systems~\cite{Scott,Meron,Goldbeter}. But there are important
differences. First of all, the excursions reported in chiral systems
are not easily visualized in the chiral monomer concentrations
themselves, but are strikingly manifested by the chiral polarization
or enantiomeric excess. Secondly, the total enantiomer and
heterodimer concentrations do increase with time, so that the
complete phase-space trajectory does not follow a closed path: there
is no return to the initial state. The chiral excursion is a one way
trip, not a round trip as in an excitable system. Third, whereas
excursions are traditionally studied for open excitable
systems~\cite{Scott,Meron,Goldbeter}, chiral excursions are observed
here only for closed or at most semi-open systems, but not for open
systems.

The original impetus for considering phase-space descriptions of the
Frank model comes not only from the chiral excursions reported
in~\cite{CHMR} and~\cite{BH} but also by the recent report of damped
chiral oscillations detected numerically in a model of chiral
polymerization in closed systems~\cite{BH}. Absolute asymmetric
synthesis is achieved in the latter scheme, accompanied by long
duration chiral excursions in the enantiomeric excesses for all the
homopolymer chains formed, analogous to the much simpler Frank model.
But unlike the latter, strong enantiomeric inhibition converts these
excursions into long period damped chiral oscillations in the
enantiomeric excesses associated with the longest homochiral polymer
chains formed.  Moreover, short period sustained chiral oscillations
have been observed numerically in a recycled Frank model open to
energy flow, for large values of the inhibition~\cite{Plasson}. This
oscillatory behavior poses an additional problem for the origin of
biological homochirality, since any memory of the sign of the initial
fluctuation is further erased by subsequent oscillations thus adding a
further element of uncertainty to the overall problem. Chemical
oscillations have been traditionally studied in conjunction with
excitability.  Although the latter concept is not directly applicable
to models exhibiting SMSB, it remains to be seen if the techniques
used to study oscillations can be applied profitably to reaction
schemes that lead to chiral oscillations.

\section*{Acknowledgments}
DH acknowledges the Grant AYA2009-13920-C02-01 from the Ministerio
de Ciencia e Innovaci\'{o}n (Spain) and forms part of the ESF COST
Action CM07030: \textit{Systems Chemistry}.  MS acknowledges support
from the Spanish MICINN through project FIS2008-05273 and from the
Comunidad Aut\'onoma de Madrid, project MODELICO (S2009/ESP-1691).
CB acknowledges a Calvo-Rod\'{e}s scholarship from the Instituto
Nacional de T\'{e}cnica Aeroespacial (INTA).

%%%%%%%%%%%%%%%%%%%%%%%%%%%%%%%%%%%%%%%%%%%%%%%%%%%%%%
%% Bibliography.
%%%%%%%%%%%%%%%%%%%%%%%%%%%%%%%%%%%%%%%%%%%%%%%%%%%%%%
%\bibliographystyle{elsart-num}
%\bibliography{main}

%

\end{document}